\documentclass[12pt]{article}
\begin{document}
\title{What if Superluminal Neutrinos Exist but not Higgs Bosons?}
\author{B.G. Sidharth\\
International Institute for Applicable Mathematics \& Information Sciences\\
Hyderabad (India) \& Udine (Italy)\\
B.M. Birla Science Centre, Adarsh Nagar, Hyderabad - 500 063
(India)}
\date{}
\maketitle
\begin{center}
{\bf (Based on the Bhagawantham Memorial Lecture 2011)}
\end{center}
\begin{abstract}
As 2011 ended, two results stood out which seemed to go against
twentieth century ideas. The first was the OPERA superluminal
neutrino observation contradicting Special Relativity. The second
was lack of a definitive appearance of the Higgs Boson. While both
these hopefully will be decided by the end of 2012, we investigate a
single mechanism that explains both these anomalies.
\end{abstract}
\section{Introduction}
One of the pillars of twentieth century physics has been Einstein's
Special Theory of Relativity according to which the speed of light
is the maximum limit in the universe.\\
Similarly another pillar of last century's physics has been the
Standard Model of Particle Physics which has been in place from
around 1970. The year 2011 saw doubts cast on both these well
established theories. First came an announcement on 23rd September
that neutrinos which were let off from CERN in Geneva, reached the
GRAN SASSO Lab in Central Italy some sixty nano seconds too early,
thus apparently breaching the speed of light barrier and the Special
Theory of Relativity. This was a $6 \sigma$ result. Even so, the
experiment was then repeated, with smaller bunches of neutrinos to
avoid possible errors. Again the same result was obtained till a major error was spotted.\\
On the other hand, after several false rumours that the Higgs Boson,
a missing but vital piece of the Standard Model had been detected at
LHC in CERN, a Conference was held on 13th December to unveil the
latest conclusions of the CMS and ATLAS teams. It was widely
anticipated that the discovery of the Higgs Boson would be
announced. Contrary to the build up of expectations, the
announcement was bland. There were mere hints of the Higgs Boson,
but no definite sighting. The definitive information that came out
was that the Higgs Boson, if it exists, would have a mass of about
$125 GeV$. Both these puzzles may be resolved by end 2012. Even so doubts are alive.\\
This prompts us to play the role of the devil's advocate and ask,
what if superluminal neutrinos are a reality, but not the Higgs
Boson? There are alternative theories that explain separately either
of these two anomalies. But we look for a single explanation for
both these apparently unrelated issues. This is to be found if we
discard a common assumption -- or plank -- on which twentieth
century physics rests: a smooth spacetime. But if spacetime were not
so, as for example in Quantum Gravity approaches, that is spacetime
were non-commutative, then both the anomalies can be explained, as
we see below.
\section{Superluminal Neutrinos}
The OPERA (Oscillation Project with Emulsion Tracking Apparatus)
experiment, $1400$ meters underground in the Gran Sasso National
Laboratory in Italy has detected neutrinos travelling faster than
the speed of light, which has been a well acknowledged speed barrier
in physics. This limit is $299792,458$ meters per second, whereas
the experiment has detected a speed of $299,798,454$ meters per
second. In this experiment neutrinos from the CERN Laboratory $730$
kilometers away in Geneva were observed. They arrived $60$ nano
seconds faster than expected, that is faster than the time allowed
by the speed of light. The experiment has been measured to $6
\sigma$ level of confidence, which makes it a certainty \cite{adam}.
However it is such an astounding discovery that the OPERA scientists
would like further confirmation from other parts of the world. In
the meantime they performed the experiment all over again, but this
time using smaller bunches of neutrinos, to eliminate certain
possible errors. The result was the same. In 2007 the MINOS
experiment near Chicago did find hints of this superluminal effect
\cite{adamson}. Nevertheless scientists wait with bated breath to confirm this earth shattering discovery.\\
The best direct test of Einstein's energy mass formula so far has
been made by combining accurate measurements of atomic mass
differences $\Delta m$ and of the $\gamma$-ray wavelengths to
determine the energy, the nuclear binding energy for isotopes of
silicon and sulphur \cite{rain}. They found that the energy mass
formula can be separately confirmed in two tests yielding a combined
result of $1 - \Delta mc^2/E = (-1.4 \pm 1.4) \times 10^{-7}$,
indicating that it holds to a level of at least $0.00004\%$.\\
It must be reported that the author had predicted such deviations
from Einstein's Theory of Relativity, starting from 2000. This work
replaces the usual Einstein energy momentum formula with the
modified expression (the so called Snyder-Sidharth Hamiltonian),
\begin{equation}
E^2 = p^2 c^2 + m^2 c^4 + \frac{\alpha c^2}{\hbar^2} l^2
p^4\label{X}
\end{equation}
where $l$ is a minimum length like the Planck length and $\alpha$ is
positive for fermions or spin half particles like neutrinos
\cite{bgscsf01,cu,bgsfpl,uof,tduniv}. The above formula is based on
considerations of a non differentiable spacetime at ultra high
energies. In this case, the usual commutative relations of Quantum
theory are replaced, as shown a long time ago by Snyder
(Cf.refs.\cite{cu,uof}) by
\begin{equation}
[x,y] = O(l^2), [x,p_x] = \imath \hbar [1 + l^2]\label{Y}
\end{equation}
Equation (\ref{X}) shows that the energy at very high energies for
fermions is greater than that given by the relativity theory so that
effectively the speed of the particle is slightly greater than that
of light. For example, if in the usual formula, we replace $c$ by $c
+ c'$, then, comparing with the above we would get:
$$c' = \alpha l^2 \cdot \frac{c^2}{\hbar^2} p^4 [4 m^2 c^3 + 2 p^2 c]^{-1}$$
The difference is slight, but as can be seen is maximum for the
lightest fermions, viz., neutrinos. The above formula simplifies
to
$$c' = \frac{\alpha}{6} c,$$
taking for example the neutrino mass to be $10^{-3}eV$ and $l$ to be
the neutrino Compton wavelength. A value, $\alpha \sim 1.2 \times
10^{-4}$ reproduces the Gran Sasso
result.\\
In any case, the author had argued in 2008 \cite{bgsfpl} that special relativity would break down at and within the Compton wavelength. For the neutrino, taking its mass as $\sim 10^{-8}$ times the electron mass, the neutrino Compton wavelength is roughly $\sim 10^{-3} cm$. Within this region, already large by particle physics standards, or equivalently within about $10^{-13}sec$., we can expect to see deviations from special relativity.\\
There are other interesting ramifications of this relation, for
example the mass of a particle and its antiparticle may differ
slightly, and so on (Cf. also \cite{nap}).
\section{The Higgs Boson}
It is well known that in the Standard Model, Peter Higgs and a few
others invoked the idea of the Higgs Boson via a BCS mechanism to
explain why and how elementary particles acquire a mass. However, as
pointed out in the introduction, from the mid sixties the search for
the Higgs has as of date failed to locate the particle. We would now
like to point out that based on the above fuzzy or non commutative
spacetime we get a formulation that mimics the Higgs mechanism to
generate mass, without actually
requiring a new particle.\\
It is well known that Hermann Weyl's original phase transformation
proposal was generalized, so that the global or constant phase of
$\lambda$ was considered to be a function of the coordinates
\cite{uof,moriyasu,jacob,greiner}.\\
As is well known this leads to a covariant gauge derivative. For
example, the transformation arising from $(x^\mu) \to (x^\mu +
dx^\mu)$,
\begin{equation}
\psi \to \psi e^{-\imath \lambda}\label{Ee1}
\end{equation}
leads to the familiar \index{electromagnetic}electromagnetic
potential gauge,
\begin{equation}
A_\mu \to A_\mu - \partial_\mu \lambda\label{Ee2}
\end{equation}
The above transformation, ofcourse, is a \index{symmetry}symmetry
transformation. In the transition from (\ref{Ee1}) to (\ref{Ee2}),
we expand the exponential,
retaining terms only to the first order in coordinate differentials.\\
Let us now consider the gauge field in some detail. As is known this
could be obtained as a generalization of the above phase function
$\lambda$ to include fields with internal degrees of freedom. For
example $\lambda$ could be replaced by $A_\mu$ given by
\cite{moriyasu}
\begin{equation}
A_\mu = \sum_{\imath} A^\imath_\mu (x)L_\imath ,\label{Eex1}
\end{equation}
The \index{gauge field}gauge field itself would be obtained by using
Stoke's Theorem and (\ref{Eex1}). This is a very well known
procedure: considering a circuit, which for simplicity we can take
to be a parallelogram of side $dx$ and $dy$ in two dimensions, we
can easily deduce the equation for the field, viz.,
\begin{equation}
F_{\mu \nu} = \partial_\mu A_\nu - \partial_\nu A_\mu - \imath q
[A_\mu , A_\nu ],\label{Eex2}
\end{equation}
$q$ being the \index{gauge field}gauge field coupling constant.\\
In (\ref{Eex2}), the second term on the right side is typical of a
non Abelian \index{gauge field}gauge field. In the case of the U(1)
\index{electromagnetic}electromagnetic field, this latter term vanishes.\\
Further as is well known, in a typical Lagrangian like
\begin{equation}
\mathit{L} = \imath \bar \psi \gamma^\mu D_\mu \psi - \frac{1}{4}
F^{\mu \nu} F_{\mu \nu} - m \bar \psi \psi\label{Eex3}
\end{equation}
$D$ denoting the Gauge \index{covariant derivative}covariant
derivative, there is no \index{mass}mass term for the field
\index{Boson}Bosons. Such a \index{mass}mass term in (\ref{Eex3})
must have the form $m^2 A^\mu A_\mu$ which unfortunately is not
Gauge invariant.\\
This was the shortcoming of the original
\index{Yang-Mills}Yang-Mills Gauge Theory: The Gauge Bosons would be
\index{mass}massless and hence the need for a \index{symmetry
breaking}symmetry breaking,
\index{mass}mass generating mechanism.\\
The well known remedy for the above situation has been to consider,
in analogy with \index{superconductivity}superconductivity theory,
an extra phase of a self coherent system (Cf.ref.\cite{moriyasu} for
a simple and elegant treatment and also refs. \cite{jacob} and
\cite{taylor}). Thus instead of the \index{gauge field}gauge field
$A_\mu$, we consider a new phase adjusted \index{gauge field}gauge
field after the \index{symmetry}symmetry is broken
\begin{equation}
W_\mu = A_\mu - \frac{1}{q} \partial_\mu \phi\label{Eex4}
\end{equation}
The field $W_\mu$ now generates the \index{mass}mass in a self
consistent manner via a Higgs mechanism. Infact the kinetic energy
term
\begin{equation}
\frac{1}{2} |D_\mu \phi |^2,\label{Eex5}
\end{equation}
where $D_\mu$ in (\ref{Eex5}) denotes the Gauge \index{covariant
derivative}, now becomes
\begin{equation}
|D_\mu \phi_0 |^2 = q^2|W_\mu |^2 |\phi_0 |^2 \, ,\label{Eex6}
\end{equation}
Equation (\ref{Eex6}) gives the \index{mass}mass in terms of the ground state $\phi_0$.\\
The whole point is as follows: The \index{symmetry breaking}symmetry
breaking of the \index{gauge field}gauge field manifests itself only
at short length scales signifying the fact that the field is
mediated by particles with large \index{mass}mass. Further the
internal \index{symmetry}symmetry space of the \index{gauge
field}gauge field is broken by an external constraint: the wave
function has an intrinsic relative phase factor which is a different
function of spacetime coordinates compared to the phase change
necessitated by the minimum coupling requirement for a free particle
with the gauge potential. This cannot be achieved for an ordinary
point like particle, but a new type of a physical system, like the
self coherent system of \index{superconductivity}superconductivity
theory now interacts with the \index{gauge field}gauge field. The
second or extra term in (\ref{Eex4}) is effectively an external
field, though (\ref{Eex6}) manifests itself only in a relatively
small spatial interval. The $\phi$ of the Higgs field in
(\ref{Eex4}), in analogy with the phase function of  \index{Cooper
pairs}Cooper pairs of \index{superconductivity}superconductivity
theory comes with a
\index{Landau-Ginzburg}Landau-Ginzburg potential $V(\phi)$.\\
Let us now consider in the \index{gauge field}gauge field
transformation, an additional phase term, $f(x)$, this being a
scalar. In the usual theory such a term can always be gauged away in
the \index{U(1)}U(1) \index{electromagnetic}electromagnetic group.
However we now consider the new situation of a
\index{noncommutative}noncommutative geometry discussed earlier
viz.,
\begin{equation}
\left[dx^\mu , dx^\nu \right] = \Theta^{\mu \nu} \beta , \beta \sim
0 (l^2)\label{Eex7}
\end{equation}
where $l$ denotes a minimum \index{spacetime}spacetime cut off.
Equation (\ref{Eex7}) is infact \index{Lorentz}Lorentz covariant.
Then the $f$ phase factor gives a contribution to the second order
in coordinate differentials,
$$\frac{1}{2} \left[\partial_\mu B_\nu - \partial_\nu B_\mu \right] \left[dx^\mu , dx^\nu \right]$$
\begin{equation}
+ \frac{1}{2} \left[\partial_\mu B_\nu + \partial_\nu B_\mu \right]
\left[dx^\mu dx^\nu + dx^\nu dx^\mu \right]\label{Eex8}
\end{equation}
where $B_\mu \equiv \partial_\mu f$.\\
As can be seen from (\ref{Eex8}) and (\ref{Eex7}), the new
contribution is in the term which contains the commutator of the
coordinate differentials, and not in the symmetric second term.
Effectively, remembering that $B_\mu$ arises from the scalar phase
factor, and not from the non-Abelian \index{gauge field}gauge field,
in equation (\ref{Eex2}) $A_\mu$ is replaced by
\begin{equation}
A_\mu \to A_\mu + B_\mu = A_\mu + \partial_\mu f\label{Eex9}
\end{equation}
Comparing (\ref{Eex9}) with (\ref{Eex4}) we can immediately see that
the effect of noncommutativity is precisely that of providing a new
\index{symmetry breaking}symmetry breaking term to the \index{gauge
field}gauge field, instead of the $\phi$ term, (Cf.refs.
\cite{cr39,ijmpe}) a term not belonging to the
\index{gauge field}gauge field itself.\\
On the other hand if we neglect in (\ref{Eex7}) terms $\sim l^2$,
then there is no extra contribution coming from (\ref{Eex8}) or
(\ref{Eex9}), so that we are in the usual non-Abelian \index{gauge
field}gauge field theory, requiring a broken
\index{symmetry}symmetry to obtain an equation like (\ref{Eex9}).
\section{Comments}
We note that some of the objections to the superluminal neutrino
experiment have included: Energy loss due to Cerenkov radiation and
consequently a slowing down of the neutrinos. Or an error in the
GPRS determination of emission and arrival times. This will be
addressed with fibre optic network instead of GPRS. On the other
hand extra dimensions of String theory have been invoked to explain
the superluminal feature.\\
As for the Higgs Bosons, effects like techni-colour have been
invoked as a mass generating mechanism.\\ \\

\begin{flushleft}
{\bf \Large{Appendix}}
\end{flushleft}

\noindent To see the above modified SS Hamiltonian in greater
detail, we note that, as shown by the author in early 2000
\cite{bgsdiscrete} given a minimum length $l$, the energy momentum
relation gets modified. The usual Quantum Mechanical commutation
relations get modified as shown in Section 2 (Cf.(\ref{Y})) and now
become
\begin{equation}
[x,p] = \hbar' = \hbar [1 + \left(\frac{l}{\hbar}\right)^2 p^2]\,
etc\label{5He2}
\end{equation}
(Cf. also ref.\cite{bgsust}). (\ref{5He2}) shows that effectively
$\hbar$ is replaced by $\hbar'$. So, in units, $\hbar = 1 = c$,
$$E = [m^2 + p^2 (1 + l^2 p^2)^{-2}]^{\frac{1}{2}}$$
or, the energy-momentum relation leading to the Klein-Gordon
Hamiltonian is given by,
\begin{equation}
E^2 = m^2 + p^2 - 2l^2 p^4,\label{5He3}
\end{equation}
neglecting higher order terms. This is the so called Snyder-Sidharth
Hamiltonian for Bosons \cite{glinka}. (It may be mentioned that some
other authors have since ad hoc taken a third power of $p$, and so
on \cite{myers}. However we should remember that these were mostly
phenomenological approaches.)\\
For Fermions the analysis can be more detailed, in terms of Wilson
lattices \cite{mont}. The free Hamiltonian now describes a
collection of harmonic fermionic oscillators in momentum space.
Assuming periodic boundary conditions in all three directions of a
cube of dimension $L^3$, the allowed momentum components are
\begin{equation}
{\bf q} \equiv \left\{q_k = \frac{2\pi}{L}v_k; k = 1,2,3 \right\},
\quad 0 \leq v_k \leq L - 1\label{4.59}
\end{equation}
(\ref{4.59}) finally leads to
\begin{equation}
E_{\bf q} = \pm \left(m^2 + \sum^{3}_{k=1} a^{-2} sin^2
q_k\right)^{1/2}\label{4.62}
\end{equation}
where $a = l$ is the length of the lattice, this being the desired
result leading to
\begin{equation}
E^2 = p^2e^2+m^2c^4 + \alpha l^2p^4\label{ex} \end{equation}
(\ref{ex}) shows that $\alpha$ is positive, that is for Fermions the
Snyder-Sidharth Hamiltonian is given by (\ref{ex}) as noted in
Section 2.\\
We point out that using the modified dispersion relation (\ref{ex}),
for a massless particle, $m = 0$, and identifying the extra term
$l^2p^4$ as being due to a mass $\delta m$, we can easily deduce
that, restoring proper units,
$$\frac{c^2}{\hbar^2} l^2 p^4 = \Delta E^2 = \delta m^2 c^4,$$
Whence,
$$\delta m = \frac{\hbar}{cl} \quad \mbox{or} \quad l =
\frac{\hbar}{c\delta m}$$ This shows that $l$ is the Compton
wavelength for this mass $\delta m$ or alternatively if $l$ is the
Compton wavelength, then we deduce the mass, now generated from the
extra effect. This is another demonstration of mass generation from
$O(l^2)$ effects as seen in Section 3, without requiring a Higgs
mechanism. If, for example, $l$ were the Planck length, then $\delta
m$ would be the Planck mass (and vice versa).

\end{document}